\documentclass{aastex}
\usepackage{graphicx}

\begin{document}

\title{A DISK OF DUST AND MOLECULAR GAS AROUND A HIGH-MASS PROTOSTAR}

\author{N. A. Patel\altaffilmark{1}, S. Curiel\altaffilmark{1,2},  T. K. 
Sridharan\altaffilmark{1}, Q. Zhang\altaffilmark{1},  T. R.  
Hunter\altaffilmark{1}}
\author{P.T.P. Ho\altaffilmark{1}, J. M. Torrelles\altaffilmark{3},  J. M. 
Moran\altaffilmark{1}, 
J. F.  G\'omez\altaffilmark{4}, G. Anglada\altaffilmark{5}}

\altaffiltext{1}{Harvard-Smithsonian Center for Astrophysics, Cambridge, MA, 
USA}
\altaffiltext{2}{Instituto de Astronomia, UNAM, Unidad Morelia, Mexico}
\altaffiltext{3}{Consejo Superior de Investigaciones Cient\'{\i}ficas-IEEC, 
Spain. On sabbatical
leave at the UK Astronomy Technology Centre, Royal Observatory Edinburgh}
\altaffiltext{4}{Laboratorio de Astrofisica Espacial y Fisica Fundamental, INTA, 
Madrid, Spain}
\altaffiltext{5}{Instituto de Astrofisica de Andalucia, CSIC, Granada, Spain}

{\bf The processes leading to the birth of low-mass stars such as
our Sun have been well studied$^{1}$, but the formation of high-mass
($> 8\times$ Sun's mass M$_{\odot}$) stars has heretofore remained
poorly understood$^{2}$. Recent observational studies suggest that
high-mass stars may form in essentially the same way as low-mass stars,
namely via an accretion process$^{3}$, instead of via
merging of several low-mass ($< 8$M$_{\odot}$) stars$^{4}$.   However,
there is as yet no conclusive evidence$^{5,6}$.  Here, we report the
discovery of a flattened disk-like structure  observed at submillimeter
wavelengths, centered on a massive 15 M$_{\odot}$ protostar in the
Cepheus-A  region. The disk, with a radius of about 330 astronomical
units (AU) and a mass of 1 to 8 M$_{\odot}$, is detected in dust
continuum as well as in molecular line emission. Its perpendicular
orientation to, and spatial coincidence with the central embedded powerful
bipolar radio jet, provides the best evidence yet
that massive stars form via disk accretion in direct analogy to the
formation of low-mass stars.}

Previously reported disk-like structures associated with
high-mass stars, which are more than twice as far away as low-mass
stars, have typical sizes of several thousands of AU. Furthermore,
in such sources, the presence of thermal jets at scales of a few
$\times10^{2}$AU has not been demonstrated, in part due
to the lack of sufficient angular resolution.   Consequently, 
it has not been possible
to identify and isolate a good example of a high-mass protostar-disk-jet
system$^{7,8,9}$.  The recently reported massive disk in M17-S01
(ref. 5), based on near-infrared observations of the disk in
silhouette against the bright background light of the ionized region
and in emission lines of carbon monoxide (CO), has subsequently
been shown to be actually a much lower mass disk (0.09 M$_{\odot}$;
ref. 6). In the former study$^{5}$, uncertainty resulted from
(1) insufficient angular resolution (8$''$), (2) observations of
molecular species that are most likely optically thick and a tracer
of low-density gas (CO and its isotopes, $^{13}$CO and C$^{18}$O),
and (3) the absence of an independent means to infer the presence of a
high-mass protostar such as bright radio continuum emission and/or
luminous water maser sources.

Cepheus-A is a well studied high-mass star-forming region in the
Cepheus OB3 complex$^{10,11}$. The bolometric luminosity of this
high-mass star-forming region is about 2.5$\times$10$^{4}$L$_{\odot}$
(ref. 12). Half of this luminosity is attributed to the HW2 object, 
the brightest radio continuum source in the field, that is considered to be a 
B0.5 spectral type protostar of 15 $M_\odot$ (refs. 13,14,15,16)
and is the most likely exciting source of a powerful 
extended bipolar molecular
outflow$^{17,18}$. 
The radio continuum flux density of about 40 mJy at
1.3 cm wavelength$^{16}$ would correspond to a flux density of about
0.8 Jy  if the source were at the distance of a typical low-mass 
star-forming region such as the Taurus molecular
cloud (160 pc).  This would be about 100 times brighter than 
typical low-mass protostars at that distance. Also associated with HW2 is a
``biconical thermal radio jet'' at the size scale of $\simeq$1$''$
($\simeq$725 AU), with the ionized gas exhibiting proper motions
with velocities $\geq$~500 km s$^{-1}$ along the axis of the
jet$^{14,19,20}$.  In addition, we estimate that the luminosity of the
masers associated with HW2(ref. 21) is L(H$_2$O) $\geq$
3$\times$10$^{-6}$~L$_{\odot}$, which is $\sim$ 10$^{3}$
to 10$^{5}$ times more luminous than water masers associated with
low-mass stars (ref. 22). These comparisons 
strongly imply that HW2 is a high-mass protostar.

Our new sub-arcsecond angular resolution Submillimeter Array
(SMA)$^{23}$ observations toward HW2 have far superior spatial
resolution and sensitivity to dust emission, when compared to existing
submillimeter observations made with single-dish telescopes or
interferometric observations made at millimeter and centimeter wavelengths.
This allowed us to directly image for the first time the compact
circumstellar disk surrounding HW2 at the scale of a few hundred
AU, and to define its physical properties.
Figure 1 shows an overlay of the dust continuum emission at 327 GHz
and the integrated intensity in the CH$_{3}$CN J=18-17 line emission.
The dust continuum emission was imaged
by selecting spectrometer channels free of line emission. The
deconvolved source radius is 330 AU ($0\rlap.{''}45$) in dust
continuum emission, and 580 AU ($0\rlap.{''}8$) in CH$_{3}$CN line
emission.  The integrated flux density from the dust emission is
2.0 Jy $\pm$ 0.1 Jy. The Gaussian-fitted positions of the dust and
CH$_{3}$CN emission peaks agree well to within a tenth of an
arcsecond.  The sizes, orientations and other characteristics are
summarized in Table 1. By examining the visibility amplitudes as a
function of baseline distance, we have confirmed that this flattened
structure is well resolved along its major axis and partially
resolved along its minor axis.  The free-free
continuum emission from the thermal jet at 1.3 and 3.6 cm wavelengths,
observed with the Very Large Array (VLA), is also shown
in Fig. 1.  The 1.3 cm wavelength continuum emission traces the inner part of the
jet, with the protostellar source most likely at its geometric 
center$^{16,19}$.    The position
angle of the thermal jet at both 1.3 and 3.6 cm wavelengths is $\sim45^{o}$. A
large-scale extended ($\sim1'$) bipolar outflow mapped in HCO$^{+}$
J=1-0 emission is centered at the position of HW2 (ref. 17) and has
the same orientation as the much smaller scale centimeter wavelength
continuum jet.  The size and morphology of the dust and gas emission
are in good agreement with each other and the elongation in both
these emissions is seen to be nearly perpendicular to, and peaking
on the biconical circumstellar thermal jet.  This strongly supports
the disk interpretation for the flattened structure seen in the dust
and CH$_{3}$CN emission.  Based on these observations of the jet
and outflow at various wavelengths, the existence of such a disk
might be expected, since outflows  in low mass stars are in fact
launched from the inner portions of the rotating disks according
to theoretical models (e.g., ref. 26).  Although these models are
for low-mass stars, our observational results suggest that outflows
in high-mass stars may be produced in a similar fashion. 

Figure 2 shows a position-velocity diagram along the major axis of
the elongated emission seen in the CH$_{3}$CN lines. From this
diagram we see a velocity-gradient of $\simeq$6 km s$^{-1}$ over
$0\rlap.{''}5$ (considering the largest velocity shift from the
central position). We interpret this velocity gradient to be due
to gravitationally bound rotational motion. The dynamical mass
enclosed within a radius r with rotational velocity $v_{r}$ is given
by $v_{r}^{2}r/G$, where $G$ is the gravitational constant.  The
observed velocity (along the line-of-sight) is $v=v_{r}$sin(i),
where i is the inclination angle of the disk  axis with respect to the 
line-of-sight.
From the mean value of the observed aspect ratio
of gas and dust disk as listed in table 1, (i.e., assuming a circular
disk) we estimate the inclination angle to be $\simeq$62$^{o}$, and
therefore a binding mass of 19$\pm$5~M$_{\odot}$. This implies that
the observed motions can be bound by the central high-mass protostar.
In addition, from the line ratios of the CH$_{3}$CN K components
we can estimate the temperature of the gas$^{24}$. We assume optically
thin emission and a single value of temperature along the line-of-sight.
From the ratios of the peak intensities of the K=3 and K=2 components,
we find that the temperature varies along the major axis of the disk from
$\sim$25 K to $\sim$75 K, with the northwest part of the disk
relatively cooler and the southeast and central parts of the disk
relatively hotter. The line width also appears to be greater
at the latter positions. With the present angular resolution we are
unable to map in detail the temperature and kinematics of the gas
in the disk but we can make rough estimates of its mass. If we assume that the dust is
optically thin, we can estimate the gas mass of the disk$^{27}$
assuming the gas-to-dust ratio of 100, gas temperature of 50 K (from
analysis of methyl cyanide lines) and the measured 900 $\mu$m
continuum flux density of 2.0 Jy. We also need to know the grain
emissivity spectral-index $\beta$ for the dust emission. We obtain
a mass estimate for the gas disk to be 1 M$_{\odot}$ for $\beta=1$
(ref. 28) and 8 M$_{\odot}$ for $\beta=2$ (ref. 29).  Some of this
mass could be in a rotationally flattened infalling envelope instead
of the disk but we note that the observed size of the disk in HW2
is in good agreement with the range of $\simeq$ 400 to 600 AU for
the centrifugal radius (the radius where disk formation is expected to occur) 
 derived from a recent fitting of flattened infalling envelope models to  the observed
spectral energy distribution of high-mass protostars$^{30}$.  All
these results support theoretical models of high-mass star-formation
via an accretion process occurring in a disk around the protostar,
accompanied by a bipolar outflow (much like low-mass stars), rather
than by models that require merging of several low-mass stars.

\begin{deluxetable}{lcccccc}
\tablecaption{Characteristics of Cepheus-A HW2 dust continuum and CH$_{3}$CN 
emission}
\tablehead{&Total flux density &major axis ('')&minor axis ('')&P. A.  
($^{o}$)&$\Delta\alpha('')$&$\Delta\delta('')$}
\startdata
Continuum&2.0 $\pm$ 0.1 Jy &0.9 &0.5&-59.2$\pm$0.6&-0.07&0.55\\
CH$_{3}$CN&122 Jy km s$^{-1}$&1.6&0.6&-56.2$\pm$0.2&-0.04&0.58\\
\enddata
\tablecomments{ Deconvolved Gaussian fitted model parameters for the Cepheus-A 
HW2 disk structure seen in continuum and CH$_{3}$CN emission at 330 GHz (J=18-
17, K=0,1,2,3). 
The last two columns are the positions of the peaks with respect to  
$\alpha(2000)=22^{h}56^{m}17.^{s}970, 
\delta(2000)=+62^{o}01^{'}48.992^{''}$. 
The uncertainties in the sizes and positions are $\sim0.''1$.}
\end{deluxetable}
\clearpage

\centerline {\bf REFERENCES}
\begin{enumerate}
\item Lada, C. J. \& Shu. F. H. The formation of sunlike stars. {\it Science} 
{\bf 248}, 564-572 (1990).
\item Stahler, S. W., Palla, F.  \& Ho, P.T.P. in Protostars and Planets IV 
(eds Mannings, V., Boss, A. P.  and Russell, S.) 327-351 (Univ. of Arizona 
Press, Tucson, 2000). 
\item McKee, C. \& Tan, J. The formation of massive stars from turbulent cores. {\it Astrophys.
J.} {\bf 585}, 850-871 (2003).
\item  Bonnell, I. A. \& Bate, M. R. Accretion in stellar clusters and the 
collisional formation of massive stars. {\it Mon. Not. R. Astron. Soc.} {\bf 
336}, 659-669 (2002). 
\item Chini, R., Hoffmeister, V., Kimeswenger, S., Nielblock, M., Nurnberger, 
D., Schmidtobreick,  \& Sterzik, M. The formation of a massive protostar through 
the disk accretion of gas.  {\it Nature} {\bf 429},  155-157 (2004).
\item Sako, S. et al. No high-mass protostars in the silhouette young stellar 
object M17-SO1.  {\it Nature}  {\bf 434}, 995-998 (2004).
\item Zhang, Q., Hunter T. R. \& Sridharan, T. K. A Rotating Disk around a High-
Mass Young Star. {\it Astrophys. J.} {\bf 505}, L151-L154 (1998).
\item Shepherd, D., Claussen, M. J. \& Kurtz,  S. E. Evidence for a Solar System-
Size Accretion Disk Around the Massive Protostar G192.16-3.82.  {\it Science} 
{\bf 292}, 1513-1518 (2001).
\item Cesaroni, R., Felli, M., Jenness, T., Neri, R., Olmi, L., Robberto, M., 
Testi, L. \& Walmsley, C.M.  Unveiling the disk-jet system in the massive 
(proto)star IRAS 20126+4104. {\it Astron. Astrophys.} {\bf 345}, 949-964 (1999).
\item Sargent, A. I. Molecular clouds and star formation. I - Observations of 
the Cepheus OB3 molecular cloud. {\it Astrophys. J.} {\bf 218}, 736-748 (1977).
\item Blitz, L. \& Lada, C. J. H2O masers near OB associations. {\it Astrophys. 
J.} {\bf  227}, 152-158 (1979). 
\item Evans, N. J., et al. Far-infrared observations of the Cepheus OB3 
molecular cloud. {\it Astrophys. J.} {\bf  244},  115-123 (1981).
\item Hughes, V. A., \& Wouterloot, J. G. A. The star-forming region in Cepheus 
A. {\it Astrophys. J.}  {\bf 276},  204-210 (1984)
\item Rodr\'{\i}guez, L. F., Garay, G., Curiel, S., Ramirez, S., Torrelles, J. 
M., Gomez, Y. \&  Velazquez, A.  Cepheus A HW2: A powerful thermal radio jet. 
{\it Astrophys. J.} {\bf 430}, L65-L68 (1994).
\item Hughes, V, Cohen, R. \& Garrington, S. High-resolution observations of 
Cepheus A.   {\it Mon. Not. R. Astron. Soc.} {\bf 272}, 469-480 (1995).
\item Torrelles, J. M., G\'omez, J. F., Rodr\'{\i}guez, L. F., 
Curiel, S., Ho, P. T. P., Garay, G. The thermal radio jet of Cepheus A HW2 and the 
water maser distribution at 0.08$''$ scale (60 AU). {\it Astrophys. J.} {\bf 
457}, L107-L111 (1996).
\item G\'omez, J. F., Sargent, A., Torrelles, J. M., Ho, P.T.P., Rodriguez, 
L.F., Canto, J. \& Garay, G. Disk and Outflow in Cepheus A-HW2: Interferometric 
SiO and HCO$^{+}$ Observations. {\it Astrophys. J.}  {\bf 514}, 287-295 (1999). 
\item Rodr\'{\i}guez, L. F., Ho P. T. P., \& Moran, J. Anisotropic mass outflow 
in Cepheus A. {\it Astrophys. J.}  {\bf 240}, L149-L152 (1980).
\item Curiel S. et al. Large proper motions in the jet of the high-mass YSO 
Cepheus A HW2.
{\it Astrophys. J.}  submitted (2005). 
\item Rodr\'{\i}guez, L.F., Torrelles, J. M., Anglada, G., Marti, J. VLA 
Observations of Brightness Enhancements moving along the Axis of the Cep A HW2 
Thermal Jet.  {\it Revista Mexicana de Astronomia y Astrofisica} {\bf 37}, 95-99 
(2001).
\item  Torrelles, J. M., G\'omez, J. F., Garay, G., Rodr\'{\i}guez, 
L. F., Curiel, S., Cohen, R.~J. \& Ho, P. T. P. 
Systems with H$_2$O maser and 1.3 centimeter continuum emission in Cepheus A.
Astrophys. J. {\bf 509}, 262-269 (1998).
\item Furuya, R. S., Kitamura, Y., Wootten, A., Claussen, M. J. \& Kawabe, R.
Water maser survey toward low-mass young stellar objects in the northern sky 
with the 
Nobeyama 45 meter telescope and the very large array. {\it Astrophys. J. Supp.}  
{\bf 144}, 
71-134 (2003).
\item Ho.  P. T. P., Moran, J. M. \& Lo, K. Y. The Submillimeter Array.  {\it 
Astrophys. J.} {\bf 616}, L1-L6 (2004).
\item Loren, R. B. \& Mundy, L. G. The methyl cyanide hot and warm cores in 
Orion - Statistical equilibrium excitation models of a symmetric-top molecule. 
{\it Astrophys. J.}  {\bf 286}, 232-251 (1984). 
\item Pankonin, V., Churchwell, E., Watson, C. \& Bieging, J.H. A methyl cyanide 
search for the earliest stages of massive protostars. {\it Astrophys. J.} {\bf 
558}, 194-203 (2001).
\item Shu F., Najita, J., Ostriker, E. \& Shang, H. Magnetocentrifugally Driven 
Flows from Young Stars and Disks. V. Asymptotic Collimation into Jets.  {\it 
Astrophys. J.} {\bf 455}, L155-L158 (1996).
\item Hildebrand, R., The determination of cloud masses and dust characteristics 
from submillimetre thermal emission. {\it Quarterly Journal of the Royal 
Astronomical Society}  {\bf 24}, 267-282 (1983).
\item Williams, S. J., Fuller, G.A. \& Sridharan, T.K. The circumstellar 
environments of high-mass protostellar objects. I. Submillimetre continuum 
emission. {\it Astron. Astrophys.}  {\bf 417}, 115-133 (2004).
\item Hunter, T., Churchwell, E., Watson, C., Cox, P., Benford, D. \& Roelfsema, 
P. 350 Micron Images of Massive Star Formation Regions.  {\it Astron. J.} {\bf 
119}, 2711-2727 (2000).
\item De Buizer, J., Osorio, M. \& Calvet, N. Observations and Modeling of the 
2-25 $\mu$m emission from high mass protostellar object candidates.  {\it 
Astrophys. J.}, in press (2005).

\end{enumerate}
\vspace{1in} 

\noindent{\bf Correspondence} should be addressed to N. A. P. (npatel@cfa.harvard.edu)
\vspace{1in}

\noindent {\bf Acknowledgement:} SC acknowledges support from DEGAPA/UNAM,  CONACyT
grants  43120$-$E and from the Submillimeter Array project. G.A., J.F.G. and J.M.T. are supported by AYA2002-00376 grant (including 
FEDER funds). G.A. acknowledges support from Junta de Andaluc\'{\i}a. 
The Submillimeter Array is a joint
project between the Smithsonian Astrophysical Observatory and the
Academia Sinica Institute of Astronomy and Astrophysics, and is
funded by the Smithsonian Institution and the Academia Sinica. 
We are grateful to the people of Hawai'ian ancestry on whose sacred
mountain we are privileged to be guests.  
\vspace{1in}

\begin{figure}[p]
   \caption {}
 \begin{center}
   \begin{tabular}{c}
 \includegraphics[width=6.5in]{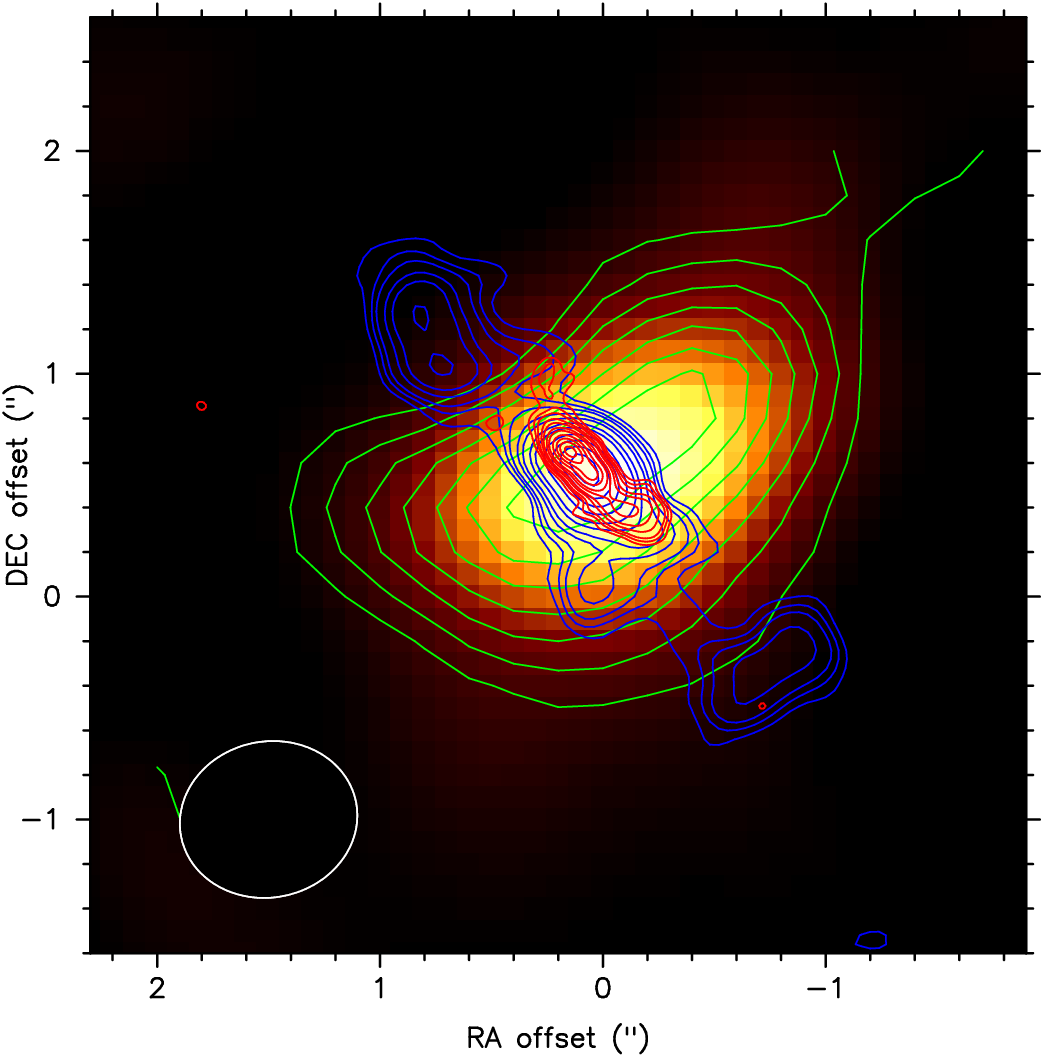}
   \end{tabular}
 \end{center}

 \end{figure}

\clearpage
\noindent{\bf Fig. 1:} Dust continuum emission of the Cepheus A HW2 protostar
at 327 GHz (halftone image ranging linearly from 0 to 1.5 Jy~beam$^{-1}$)
and integrated intensity in the CH$_{3}$CN J=18-17 (K=0,1,2,3) line
emission from -35 to 30 km s$^{-1}$ line-of-sight velocity range 
(contour levels from 5 to 40
Jy~beam$^{-1}$ km s$^{-1}$ in steps of 5 Jy~beam$^{-1}$ km s$^{-1}$).  CH$_{3}$CN
has been shown to trace high density (10$^{6}$ cm$^{-3}$) gas in
massive star-forming regions$^{24,25}$. The SMA beam size was
$0.''8\times0.''7$ with a PA of -78.$^{o}$6 (left lower corner).
This angular resolution is by far the highest reached at submillimeter
wavelength observations (e.g., compared to single-dish observations,
we have more than an order of magnitude greater angular resolution).
Blue contours show the 3.6 cm wavelength continuum emission from a  well
collimated jet, while red contours show the 1.3 cm wavelength continuum
emission from the inner part of the jet$^{16,20}$.  The protostar
is believed to be located at the origin of this elongated jet. 
The absolute astrometric error in the alignment
of the VLA and SMA observations is $\simeq 0.''2$.
The
elongation in both dust and CH$_{3}$CN  emission is nearly perpendicular
to the circumstellar thermal jet, strongly supporting the disk
interpretation.  The deconvolved disk radius is $\simeq$330 AU.
The submillimeter observations were carried out using 7 of the 8
available antennas of the SMA in the extended array configuration
with a maximum baseline of 226 meters. The observations were carried
out on 30 August 2004 and the receivers tuned to 321 GHz. With this
tuning, we had the methyl cyanide (CH$_{3}$CN) J=18-17, K=0,1,2,3...9
lines in the upper sideband and the 10$_{29}$ - 9$_{36}$ water maser
transition in the lower sideband.  Submillimeter water masers were
found to be associated with the HW2 and HW3c sources (Patel et al.
in preparation).
\clearpage

\begin{figure}
 \caption{}
 \begin{center}
   \begin{tabular}{c} \includegraphics[width=7in]{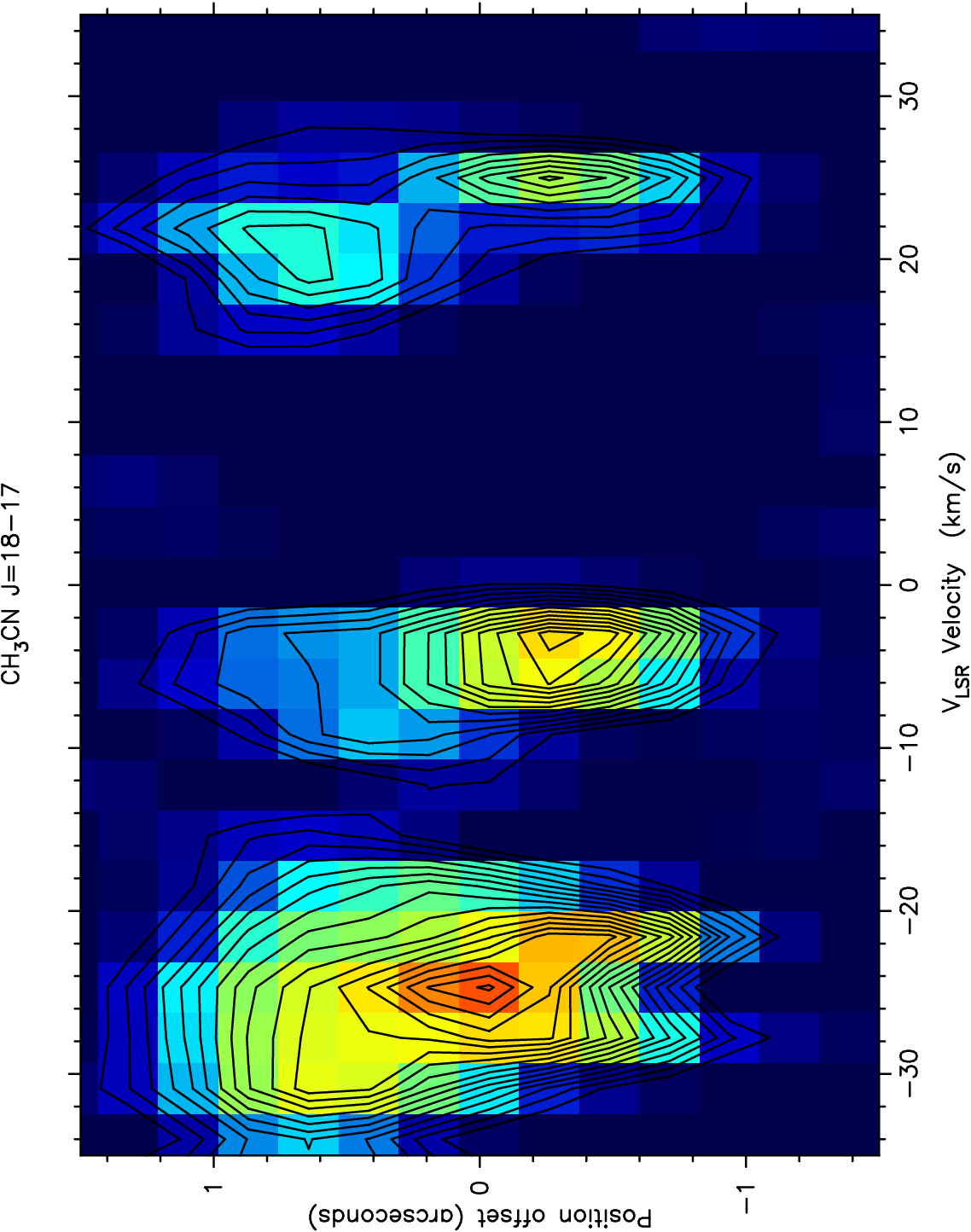}
   \end{tabular}
 \end{center}

 \end{figure}
 
 \clearpage
\noindent{\bf Fig. 2:}  Position-velocity map of CH$_{3}$CN emission along
the major axis of the elongated structure shown in figure 1 with
 position angle --53$^{o}$ (with respect to North, with East
as positive). The contour levels are from 0.3 to 1.5 Jy~beam$^{-1}$ every
0.1 Jy~beam$^{-1}$.  The position offset is measured along the major axis,
with positive offset corresponding to the southeast part and negative
offset towards the northwest part of the disk.  The 0$''$ position offset corresponds to the peak of the integrated intensity. The K = 0 and 1
components are blended. K=4 and higher lines were not detected.
The reference frequency was chosen to be at the J=18-17, K=2 line
of CH$_{3}$CN. We interpret the velocity shift of $\sim$6 km s$^{-1}$
over $0\rlap.{''}5$ seen in the K=2 and K=3 emission to be due to
rotational motion and estimate a binding mass of 19$\pm5$~M$_\odot$.
\clearpage

\end{document}